# Posture-Dependent Human $^3$He Lung Imaging in an Open Access MRI System: Initial Results

L. L. Tsai[1,2,3], R. W. Mair[1], C.-H. Li[1,4], M. S. Rosen[1,4], S. Patz[3,5]
and R. L. Walsworth[1,4]

[1] Harvard-Smithsonian Center for Astrophysics, Cambridge, MA 02138, USA.

[2] Harvard-MIT Division of Health Sciences and Technology, Cambridge, MA 02139, USA.

[3] Harvard Medical School, Boston, MA 02115, USA.

[4] Department of Physics, Harvard University, Cambridge, MA 02138, USA.

[5] Department of Radiology, Brigham And Women's Hospital, Boston, MA 02115, USA.

**Running Title:** Posture-Dependent Human $^3$He Lung Imaging

**Corresponding Author:**

Ross Mair

Harvard Smithsonian Center for Astrophysics,

60 Garden St, MS 59,

Cambridge, MA, 02138,

USA

Phone: 1-617-495 7218

Fax: 1-617-496 7690

Email: rmair@cfa.harvard.edu

# ABSTRACT

The human lung and its functions are extremely sensitive to orientation and posture, and debate continues as to the role of gravity and the surrounding anatomy in determining lung function and heterogeneity of perfusion and ventilation. However, study of these effects is difficult. The conventional high-field magnets used for most hyperpolarized $^3$He MRI of the human lung, and most other common radiological imaging modalities including PET and CT, restrict subjects to lying horizontally, minimizing most gravitational effects. In this paper, we briefly review the motivation for posture-dependent studies of human lung function, and present initial imaging results of human lungs in the supine and vertical body orientations using inhaled hyperpolarized $^3$He gas and an open-access MRI instrument. The open geometry of this MRI system features a "walk-in" capability that permits subjects to be imaged in vertical and horizontal positions, and potentially allows for complete rotation of the orientation of the imaging subject in a two-dimensional plane. Initial results include two-dimensional lung images acquired with ~ 4 mm in-plane resolution and three-dimensional images with ~ 1.5 cm slice thickness. Effects of posture variation are observed.

# INTRODUCTION

The effects of body orientation and posture changes on the regional distribution of pulmonary perfusion and ventilation have been a source of renewed interest in recent years [1-4], principally due to significant questions relating to the care and survival of patients with obstructive or restrictive lung diseases such as acute respiratory distress syndrome (ARDS) [3]. Perfusion heterogeneity has classically been attributed to effects of gravity on pleural pressure and alveolar expansion, and consequently results in regional variations in lung function [5,6]. Position-dependent changes in ventilation dynamics also play an important role in a wide variety of common clinical problems, from mechanical ventilation of the surgical patient [7] to management of respiratory complications in obese individuals [8,9]. Few methods exist that allow detailed studies of regional lung function under varying gravitational conditions – or subject orientations. Thus, pulmonary physiology could benefit greatly from the development of minimally-invasive methods to quantify regional lung function in subjects at variable orientations.

MRI has only recently been recognized as a useful tool for pulmonary imaging. Chest radiography [10] offers rapid, low-cost, high-resolution projection images with multiple subject orientations, but yields no quantitative information on gas exchange. Scintigraphy [11] offers tomographic and quantitative information but uses relatively high and costly doses of nuclear tracers and suffers from poor resolution. CT provides offers superior anatomic detail with limited functional data [12,13]. Positron emission tomography (PET) and PET/CT are used to directly measure pulmonary ventilation and perfusion and have provided the best regional quantitative detail thus far [3], but subjects are restricted to prone or supine orientations.

In recent years, MRI of inhaled, hyperpolarized $^3$He gas [14,15] has emerged as a powerful method for studying lung structure and function [16,17]. This technique is used with conventional clinical MRI instruments to make quantitative maps of human ventilation [18,19], obtain regional acinar structural information via measurements of the $^3$He Apparent Diffusion Coefficient (ADC) [20,21], and to monitor the regional alveolar gas-space $O_2$ concentration ($p_AO_2$) via the $^3$He spin-relaxation rate [22,23]. These techniques have applications to basic pulmonary physiology [24] as well as lung diseases such as asthma [25, 26], emphysema [20,27,28], lung cancer [27], and cystic fibrosis [29].

However, the large superconducting magnets used in conventional clinical MRI systems also restrict human subjects to lying in a horizontal orientation. Some initial studies with hyperpolarized $^3$He have shown that posture changes, even while horizontal, affect the lung structure modestly in a way that can nonetheless be clearly probed by $^3$He MRI [25,30,31]. However only minimal subject reorientation is possible inside conventional MRI scanners. An open-access MRI system that allows for different body orientations and postures has been developed for back lumbar studies, but is even heavier and more costly than a traditional clinical MRI scanner [32-34]. Also, the size, weight and technical restrictions of traditional clinical MRI systems demand patients be brought to the scanner. For many critical-care patients the requirement of being moved from the Intensive Care Unit is dangerous, time consuming, and expensive. Thus, the potential medical benefits of hyperpolarized $^3$He MRI are not realized for many of the most needy patients. An open-access, light-weight and less-cumbersome MRI system, therefore, could have significant potential for monitoring critically ill lung patients.

To enable posture-dependent lung imaging, we developed an open-access MRI system based on a simple electromagnet that operates at a field strength approximately 200 times lower than a traditional clinical MRI scanner. To perform MRI at such a field strength, we exploit the practicality of hyperpolarized $^3$He MRI at magnetic fields < 10 mT [35-38]. $^3$He hyperpolarized to 30–60% can be created by one of two laser-based optical pumping processes [14,15] prior to the MRI procedure; and then high-resolution gas space imaging can be performed without the need of a large applied magnetic field. Such high spin polarization gives $^3$He gas a magnetization density similar to that of water in ~ 1 T fields, despite the drastically lower spin density of the gas. Thus the signal-to-noise ratio (SNR) of hyperpolarized noble gas MRI in animal or human lungs is only weakly dependent on the applied magnetic field [33], and very-low-field MRI becomes practical [36,37]. In addition, once the effects of reduced magnetic susceptibility-induced background gradients and the resultant much longer $^3$He $T_2$* time at very-low fields are accounted for, it can be shown that the optimum field strength for hyperpolarized $^3$He may be around 0.1 T, not 1.5 or 3.0 T [39]. In addition, operation at ~ 10 mT should provide image SNR within a factor of 2–4 of that obtained in clinical scanners [39]. Other groups have recognized the benefits of low-field MRI for human studies with hyperpolarized gases [40-44], however, these studies generally have employed horizontal bore MRI magnets which restrict the subject to a single orientation [40-42,44]. One study employed a vertically-oriented electromagnet that allowed the subject to stand vertically, but not be imaged horizontally [43].

In this paper, we present initial results on posture-dependent $^3$He human lung imaging obtained with our open-access MRI system. With this system, the subject is unrestricted by the magnet and gradient coils in two dimensions. The system allows for complete re-orientation of the subject into any inclined, recumbent or inverted posture in a two-dimensional plane. Initial two- and three-dimensional human lung images are presented from subjects in two orientations - lying horizontally (supine) and sitting vertically.

# BACKGROUND

Development of an open-access MRI scanner for posture-dependent human lung imaging is motivated by: (i) current interest in the effects of body orientation and posture changes on the regional distribution of pulmonary perfusion and ventilation; and (ii) the lack of existing open-access imaging technology that does not employ ionizing radiation but does allow visualization and functional mapping of the lung in different orientations.

Regional heterogeneity of pulmonary ventilation and pulmonary perfusion is well-known to be influenced by gravity [4,45], but is also affected by the lung parenchyma and surrounding organs and stroma, leading to controversy over which effect is more physiologically relevant [5,6]. Pulmonary functional residual capacity (FRC) and gas elimination has been shown to be gravity-dependent [46], suggesting differences in local airway resistance. This has important clinical implications in mechanical ventilation, for example, where patients who are ventilated in a prone position tend to have improved gas exchange compared to those lying supine [1,2]. Non-horizontal orientations, such as Trendelenburg posturing, have been attributed to increased total respiratory elastance and resistance, mainly due to decreasing lung volumes [7]. The reverse-Trendelenburg posture is commonly used in abdominal laproscopic surgeries, where insufflation of the abdominal cavity is well known to have global cardiopulmonary effects due to increased intraperitoneal pressure [47].

Of particular interest is the change in gas exchange dynamics when a subject is moved from a supine to an upright position. This change in orientation displaces abdominal contents inferiorly, lowering the diaphragm and distending the lungs. Ribcage motion during the breathing cycle is also increased as a result of the altered load as well as changes in respiratory muscle tone [48].

Total FRC and conductance is known to significantly decrease in normal individuals undergoing such a postural change [49,50], but regional dynamics have not been measured. Although this has a small net effect on the overall respiratory mechanics of normal individuals, it can have a profound impact in disease. For example, in obese patients the total lung capacity (TLC) and FRC are markedly decreased in either posture, while baseline airway resistance is increased and maximized in the supine state [8]. Although it is generally understood that global lung mechanics are altered due to the increased load from surrounding abdominal contents and subcutaneous fat in such individuals, this provides an inadequate explanation for the observed changes in pulmonary function [9]. Of similar interest is the effect of pregnancy on pulmonary dynamics, which involves not only mass effect and mechanical changes to the cardiopulmonary circuit, but also fundamental changes to neuro-respiratory drive of the mother [51]. Management of such issues becomes increasingly relevant in modern medicine as the prevalence of obesity and asthma increases within the pregnant population and advances in fertility treatments allow for pregnancy at increasingly advanced maternal ages.

Regional measurements reflecting local gas dynamics and airway conductance are necessary to locate areas where the most significant physiological changes are occurring within the lung. To date, all pulmonary function tests performed on upright individuals have been via spirometry, with global resistance/conductance measurements limited to forced oscillation techniques performed at the mouth only. $p_AO_2$, ADC, and ventilation-perfusion ratio ($V/Q$) maps of the lung, obtained either through MRI [22,23,52] or PET [53-55] imaging, are capable of reliably resolving regional dynamics and anatomical features, but all studies have been performed on supine or prone individuals only.

The open-access magnet design of our MRI system allows for $^3$He lung imaging of subjects in either recumbent or upright postures. The ability to image both postures within the same system offers two major advantages: (i) comparative studies between the supine and upright state can be performed on the same instrument, eliminating a potentially major source of systematic error; and (ii) supine imaging with the system can be compared to the numerous supine studies already performed with PET or MRI, serving as a calibration and verification tool for the measurement techniques employed for both supine and upright lung imaging. Any gravity-dependent effects on regional lung ventilation and gas exchange seen in studies of prone and supine subjects would be further enhanced in the upright orientation, as the total vertical distance occupied by the lungs would be increased by at least 10 cm. $^3$He Ventilation images and $p_AO_2$ and ADC maps of the upright lung will provide previously unobtainable data pertaining to normal human lung physiology in a common, natural posture. For example, qualitative ventilation maps or quantitative measurements of $^3$He gas clearance over multiple breath cycles, in upright versus supine postures, will highlight regions of significant change in residual volume and resistance. These variations are likely to correlate with subject body mass index (BMI), even within normal ranges, due to mechanical and physiological changes mentioned previously. Quantification of these changes would be a significant step towards future studies involving obese patients. The open-access design of our imager is also well-suited to image individuals with extreme morbid obesity, whose size often excludes them from imaging via conventional PET or MRI systems.

Another important application of the open-access lung imager will be $^3$He MRI of asthmatic subjects. To date, $^3$He MRI with traditional clinical scanners have shown profound ventilation defects in supine subjects, even at an asymptomatic state [25]. These defects are both transient and mobile throughout the periphery of the lung. Although asthma is classically viewed as a disease of the major central airways where hypersensitive smooth muscles lead to obstructions in airflow, there is increasing evidence that peripheral lung involvement is also important [56-60]. Supine imaging with PET has shown that ventilation defects are composed of clusters of constricted terminal bronchioles; this supports a lung branching model which possesses an intrinsic sensitivity to minor instabilities, leading to major regional patches of airway collapse [61]. Importantly, such ventilation defects have usually been observed in gravity-dependent regions of the lung, so imaging in the upright posture may reveal a more exaggerated or altered distribution pattern. $^3$He ventilation, $p_AO_2$, and ADC studies on upright healthy, non-asthmatic subjects will also provide useful information, as similar, smaller-scale ventilation defects have been seen even in healthy lungs [55]. Results from these studies, as well as future studies with asthmatics, will be useful not only to further understand the mechanics of airway constriction in diseases such as asthma, but also to help understand how and why inhaled therapeutics, usually administered with the patient in an upright posture, can be effective for some patients but not for others, and how co-existing conditions such as obesity can complicate management.

The quality-of-life amongst the ever-growing geriatric population is often reduced by dyspnea, which generally results from age-related changes in pulmonary function. Anatomical and physiological changes in older lungs include smaller airway size, increased peripheral airway resistance leading to air trapping, and increasing pulmonary compliance, characteristics that can be emphysema-like [62]. Some of these alterations can be present by the age of forty, even among healthy individuals. Although TLC remains fairly constant with age, inspiratory and expiratory residual volumes increase while tidal volume decreases, exacerbating regional air trapping in the periphery [63,64]. $^3$He ADC, $p_AO_2$, and ventilation imaging of healthy adults with an open-access MRI system is expected to reveal age-dependent regional changes in airway dimensions and ventilation defects from peripheral bronchiole airway closure and may show an association between some of those changes and subject age. Equally important is the study of how and to what extent these changes are influenced by postural variations — an investigation that would be possible with an open-access MRI system. Such data would help elucidate, for example, how some healthy elderly patients who do not have classical sleep apnea can experience respiratory discomfort while lying supine but significantly alleviate their symptoms in a slight reverse-Trendelenburg orientation. This study may also be influential in the ICU setting, where mechanical ventilation in geriatric patients is often associated with poorer outcomes [65]. Potential studies involving geriatric patients would be further aided by an open-access MR system operating at very-low magnetic field, which would allow easier access than a traditional clinical scanner for disabled patients, as well as those with metallic or electronic implants who would be excluded from traditional clinical MRI systems.

Finally, idiopathic pulmonary hypertension (IPH) is a public health problem of increasing interest [66]. This disease is typically discovered around the fourth decade, after complaints of dyspnea at rest or with minor activity. However, *V/Q* mismatches can manifest themselves symptomatically under heavy exercise stress, e.g., in high-performance athletes, up to two decades earlier. In the case of chronic thromboembolic pulmonary hypertension (CTPH) [67], subjects suffer multiple microscopic, radiographically-invisible pulmonary embolisms that can eventually lead to clinical symptoms of pulmonary hypertension. The initial clinical picture can mimic early IPH, where symptoms become apparent only with major exercise. Suspicion of CTPH requires a stress exercise test where patients are monitored invasively with pulmonary arterial catheters and real-time blood gas analysis [68]. An open-access MRI system, which would permit patients to undergo an exercise challenge while inside the MRI scanner, and then be monitored in a non-invasive manner, including local measurements of $O_2$ distribution and consumption, would be a significant improvement in the study of the progression of this disease.

## EXPERIMENTAL

*Imager Design*: A detailed description of the design and operation of our open-access human MRI system will be presented elsewhere [69]. Here, a brief overview of the design, and how it permits open-access and variable-posture lung imaging, is provided.

The imager operates at an applied static magnetic field, $B_0$, of 6.5 mT (65 G). The $B_0$ field is created by a four-coil, bi-planar magnet design [70] with pairs of coils measuring 2 and 0.55 meters in diameter. One large and small coil are mounted together on a 2.2 m-diameter aluminum flange. There are two of these flange and coil sets, which are wound and arranged in a mirroring fashion, and are mounted vertically on a customized stand made of extruded aluminum beams, maintaining a separation of ~ 90 cm between coil sets. All four coils are connected in series to a single DC power supply that supplies 42.2 A of current to reach the desired $B_0$ of 6.5 mT. This field allows $^3$He MRI at a frequency of 210 kHz. With manual shimming and DC offsets on the gradient coils, the magnetic field exhibits a total variation of less than ~ 5 µT (0.05 G) across the volume of a human chest, which allows $^3$He NMR signals from such volumes with spectral FWHM line-widths of ~ 30 Hz.

Planar gradient coils were built to provide the pulsed magnetic field gradients, thus eliminating another restrictive cylindrical geometry found in clinical MRI scanners. The coils were designed to allow the acquisition of 256 × 256 $^3$He images across a 40 cm field of view (FOV) with an imaging bandwidth of 10 kHz, while avoiding concomitant field effects [71]. We wound the coils using insulated magnet wire on non-conducting, free-standing frames to facilitate heat dissipation, and minimize eddy current formation in the $B_0$ coil/flange structure. The $z$ coils consist of two sets of three circular loops, with each set mounted on the magnet flange, parallel to the $B_0$ coils. The $x$ and $y$ gradients consist of free-standing rectangular grids mounted on each magnet flange, and maintaining the ~ 90 cm spacing for subject access. The gradient coils are powered by Techron 8607 gradient amplifiers, operating at up to 140 A. At maximum current, the three gradients each provide ~ 0.07 G/cm gradient strength.

RF-frequency and gradient control is accomplished using a Tecmag Apollo commercial MRI research console [Tecmag Inc, Houston, TX]. This system is designed to operate at frequencies as low as 100 kHz without further hardware modification, unlike traditional MRI scanners. RF pulses from the Apollo are fed to an NMR Plus 5LF300S amplifier [Communications Power Corp. Inc., Hauppage, NY] which provides up to 300 W of RF power. We employed a single RF coil for $B_1$ transmission and detection, in conjunction with a Transcoupler II probe interface-T/R switch [Tecmag] optimized for 200 kHz operation. The RF coil is a large solenoid designed to accommodate the subject's shoulders and arms, and completely cover the thoracic region. The coil is ~ 50 cm in diameter and length, and is tuned to 210 kHz using an external capacitive resonance box that is remote from the coil. Being a solenoid, the coil has very high $B_1$ homogeneity [69], and can be rotated along with the subject in the imaging plane, while remaining perpendicular to the direction of $B_0$. The coil has a quality factor $Q$ ~ 30, implying operating bandwidths of ~ 10 kHz at the $^3$He Larmor frequency of 210 kHz. This low coil $Q$ removes the effect of the coil response function being convolved with the image dataset, as we had seen previously with coils of higher $Q$ and lower Larmor frequencies [38]. At a Larmor frequency of 210 kHz, the human body does not effect the coil $Q$ and has minimal loading effects, allowing the RF coil power and flip angles to be calibrated ahead of time and remain reproducible from subject to subject, unlike the case for operation at > 10 MHz frequencies in traditional clinical MRI scanners.

To improve SNR, the $B_0$, gradient and $B_1$ coils were housed inside an RF shielded room [Lindgren RF Enclosures Inc., Glendale Heights, IL]. The room attenuates environmental RF interference in the range 10 kHz to 10 MHz by up to 100 dB. Power lines for the $B_0$ magnet, preamplifier and RF coil connections all pass through commercial filters that shield out noise above 10 kHz [Lindgren Inc.]. The gradient lines pass into the shielded room via three sets of custom high-current passive line filters that produce ~ 25 dB attenuation at 100 kHz [Schaffner Inc., Luterbach, Switzerland]. The complete imager system, demonstrating subject access, is shown in Fig. 1.

*MRI Techniques*: We employed standard two- and three-dimensional fast gradient-recalled echo (FLASH) sequences for image acquisition. To maximize the non-renewable magnetization from hyperpolarized $^3$He, low-flip angle excitation pulses were used throughout. 2D projection images were acquired without slice selection, using an excitation flip angle of ~ 5º, data-set size of 128 × 64 points, 50 × 50 cm field of view (FOV) in ~ 5 seconds. 3D images employed a third-dimension phase encoding gradient that yielded a 3D dataset of size 128 × 64 × 6 across a FOV of 50 × 50 × 12 cm, using an excitation flip angle of ~ 4º, acquired in ~ 30 seconds. All imaging acquisitions used the following parameters: spectral width = 4.0 kHz, 2.2 ms sinc-shaped RF pulse, TE/TR ~ 29/86 ms, NEX = 1. The datasets were zero-filled to 128 × 128 (2D) or 128 × 128 × 8 (3D) points before fast-Fourier-transformation.

*Polarized $^3$He Production and Delivery*: Hyperpolarized $^3$He gas is produced via the spin-exchange optical pumping technique using vaporized Rb as an intermediate [14]. Our modular

$^3$He polarization apparatus includes gas storage, transport, and delivery stages [37]; recent modifications are described here. The polarization cells are ~ 80 cm$^3$ in volume, and made of Pyrex glass. A magnetic field of ~ 2.3 mT is generated by a 5-coil arrangement mounted on the polarizer, thereby providing a quantization axis for optical pumping. The 2.3 mT field also allows in-situ polarization monitoring via NMR detection at a Larmor frequency of 74 kHz using a benchtop Aurora spectrometer [Magritek, Wellington, New Zealand]. The polarizer is located adjacent to, but outside, the RF shielded room. For each experiment, we filled a polarization cell with ~ 5 - 6 bar of $^3$He and 0.1 bar of $N_2$, heated the cell to ~ 170°C, and applied ~ 30 W of circularly polarized light at 794.7 nm, provided by a line-narrowed diode laser apparatus that has an intense spectral output at 794.7 ± 0.1 nm [Spectra Physics Inc., Tuscon, AZ].

After spin-exchange optical pumping for ~ 8 – 10 hours, the $^3$He nuclear spin polarization reaches ~ 20 – 40 %. We then expand the polarized gas from the optical pumping cell into a previously evacuated glass and Teflon compressor for storage and delivery. The polarized $^3$He is then delivered via Teflon tubing through a feedthrough in the RF shielded room to a delivery manifold adjacent to the subject. This manifold consists of a Tedlar bag, vacuum and inert gas ports, and a Teflon tube through which the gas is inhaled. The valves on the manifold are controlled pneumatically.

*Human Imaging Protocol*: Figure 2 shows subjects in the open-access imager, in both horizontal and vertical orientations. After a relaxed expiration, the subjects inhale, through the Teflon tubing, ~ 500 cm$^3$ of hyperpolarized $^3$He gas, usually followed by a small breath of air to wash the helium out of the large airways and distribute it throughout the lung. The MR imaging sequence begins immediately after inhalation, and proceeds while the subject maintains a breath-hold for ~ 30 – 40 seconds. We monitor the subject's heart rate, blood pressure and blood oxygen saturation ($S_pO_2$) throughout. Subjects were restricted to healthy adults between 18 and 60 years of age, with BMI < 30, a resting $S_pO_2$ > 95%, and no history of pulmonary or cardiological disease. All human experiments are performed according to a protocol approved by the Partners Human Research Committee at Brigham and Women's Hospital/Massachusetts General Hospital, under an inter-institutional IRB agreement with the Harvard University Committee for the Use of Human Subjects in Research.

## RESULTS AND DISCUSSION

As can be seen in the photographs in Figure 2, the imager easily accommodates subjects in the supine and vertical orientations without being significantly encumbered or having their posture influenced by the RF coil. Figure 3 shows example two-dimensional human $^3$He MRI lung images, acquired without slice selection in the open-access human MRI system, operating at a field strength of 6.5 mT. Both images have a coronal orientation, with the lungs viewed in an anterioposterior direction, i.e., the subject's right lung is on the left side of the images. The images exhibit both an absence of artifacts and high signal-to-noise ratio (SNR). SNR varies from ~ 25 – 80 (Figure 3a) and ~ 50 – 140 (Figure 3b). The images were obtained without using a variable flip angle for reproducible transverse magnetization from each successive RF pulse [72], as the excitation flip angle was sufficiently low to ensure minimal variation in magnetization over the first ~ 50 phase-encoding rows, and hence produce an artifact-free image. The in-plane resolution in both images is ~ 4 mm.

The horizontal image (Figure 3a) shows the two lungs with the usual concave shape at the bottom as the diaphragm pushes against the lungs. As this image did not employ slice selection, the boundary is not sharp, but rather shows that the diaphragm impacts the front portion of the lower lung, while the lungs extend down below the diaphragm at the back. A region of lower intensity in the middle region of the left lung is consistent with the location of the heart and major aorta. The gas distribution is uniform throughout the two lungs, as expected for a healthy subject with low body mass in this orientation. The vertical image (Figure 3b) shows modest distension of the lungs in this same subject. Also, in the vertical orientation, the effect of the diaphragm on the lower portion of the lung is clearly absent. The helium gas distribution remains highly uniform, and a small trace of gas can be observed in the trachea and major bronchi.

The main application of the open-access imager is posture-dependent functional lung studies, e.g.; $p_AO_2$ mapping, which relies on accurately measuring MRI signal attenuation as a function of time. In such measurements, the use of narrow slice-selective imaging methods can lead to reduced accuracy in quantitative data as out-of-slice magnetization can diffuse into the image slice during pulse-sequence and inter-image delays and so result in an apparent signal attenuation that is lower than would be expected due to the measurement alone [73-75]. This can result in underestimations of $p_AO_2$ by as much as factor of four [75]. We note that early $p_AO_2$ mapping was also performed using projection images, partly for this reason [22,52], while recent studies incorporating slice selection have used only one or a few slices [23,76], followed by binning of multiple pixels together for analysis. In addition to avoiding thin-slice selection methods for $p_AO_2$ mapping, multi-slice experiments become more difficult to implement at the much lower RF frequencies used in the open-access imager than at the higher frequencies used in traditional clinical scanners. Generally, slices away from the magnet iso-center are achieved by varying the frequency of the slice-selective RF pulse by an amount proportional to the distance of the center of the slice from the magnet center. This frequency offset is usually in the range of ~ 1 – 10 kHz.

At a Larmor frequency of 210 kHz, and with a coil $Q$ of ~ 30, the coil response has a frequency width of ~ 7 kHz. Therefore, frequency offsets of anything more than a couple of kHz would significantly attenuate the RF pulse received by the sample, and any slice selection away from the magnet iso-center would result in variable RF calibrations.

For the reasons above, we have not implemented slice-selection in 2D imaging methods, and have instead used third-dimension phase encoding to spatially resolve the lung in the third dimension. Figures 4 and 5 show two example three-dimensional lung image datasets acquired with the open-access imager. Figure 4 was acquired while the subject was horizontal, in the supine position, while Figure 5 was acquired with a different subject sitting vertically. Both figures show multiple image planes of ~ 1.5 cm thickness, displaying the images in the anterio-posterior view, and from anterior to posterior in the image montage (# 1 – 8).

The 2D image planes of Figure 4 define the edges of the lungs, and show an even distribution of $^3$He throughout the periphery. The regions exhibiting the greatest anterio-posterior thickness correspond to the regions of greatest intensity levels from the 2D projection image. Anterior planes # 2 – 4 show the cardiac cavity, and also illustrate the characteristic concave curvature from the diaphragm, which is absent from the posterior planes # 6 – 8. The central planes also show a faint $^3$He signal in the left bronchus, but not in the right one. We attribute this to residual $^3$He remaining in the large airways even after the subject had taken a chaser breath of room air. This is consistent with anatomy; the left bronchus has a slightly sharper branching angle, and thus gas flow through the region is lower in comparison to the right bronchus. Despite the use of the third-dimension phase encoding, the image SNR in each plane remains high, generally around ~ 40 – 60 for planes # 3 – 6, and ~ 15 – 30 for the peripheral planes. For the 3D image of figure 5, the subject did not take an additional breath of room following their inhalation of the supplied helium. As a result, the 2D image planes show intense signal from $^3$He gas in the oral cavity and upper airways, while the gas is not uniformly distributed in the lungs

We emphasize two significant benefits of performing pulmonary MRI at low $B_0$ and Larmor frequency. Firstly, we operate well below the frequency range in which "sample noise" dominates human MRI [36,38]. Thus, we found that placing the RF coil over the subject resulted in minimal coil loading effects in comparison to an empty coil. The coil $Q$ was not affected in any way by the presence of the subjects, while the coil resonance moved a very small and reproducible amount [69]. In addition to making sample noise insignificant, this effect eliminates tuning/matching errors and variation in RF pulse calibrations from subject to subject. Secondly, we can use an open, bi-planar electromagnet to create a horizontal $B_0$, and thereby allow a solenoid RF coil to be used, which can then be rotated with the subject through various postures (See Figure 2). In addition to being the most sensitive RF detectors, solenoidal coils have the most homogeneous $B_1$ field among common RF coil designs [77]. Therefore, a significant confounding step in pulmonary functional imaging methods such as $p_AO_2$ mapping: i.e., the need to calibrate the effect of RF flip angle effects on a pixel-by-pixel basis for every trial with every subject [22,23,52], is not necessary in our open-access imager. Instead, it is possible to calibrate the RF coil pulse power in advance using appropriate phantoms [69], and apply this known value to later image post-processing procedures.

Finally, we note that the images presented here were acquired with a readout gradient of ~ 0.7 mT/m (0.07 G/cm), which is an order of magnitude lower than values used in traditional clinical MRI scanners. At large $B_0$, gradients of ~ 10 mT/m are employed to enable rapid echo acquisition (~ 5 – 10 ms per row), and, specifically for pulmonary imaging, to ensure that the pulsed gradient fields dominate the susceptibility-induced background gradient fields in the human lung, which scale with $B_0$. When operating at much lower $B_0$, the need for large readout gradients to dominate background gradients is no longer relevant. Additionally, use of low gradient strengths ensure that the maximum gradient across the sample, and hence gradient deviation from linearity, remains low in comparison to $B_0$, avoiding concomitant field effects [71]. However, when using such low-strength encoding gradients, correspondingly longer echo acquisition and repetition times must be used in order to achieve image resolution comparable to that obtained with traditional clinical MRI scanners, which reduces image temporal resolution. A straight-forward step to increase gradient strength, and so improve image temporal resolution, is to operate the gradient current amplifiers in series to double the maximum current available to the gradient coils. The design of the coils is not limited to a given current, and will support significantly higher currents than those used in the system's present configuration. Modification to the design of the gradient coils could also double the gradient strength. Such steps could improve temporal resolution by a factor of 4, while still ensuring the overall field gradient is low enough to avoid image distortion due to concomitant field effects [71].

## CONCLUSION

We have demonstrated human lung imaging in both the horizontal and vertical body orientations using inhaled hyperpolarized $^3$He gas and an open-access MRI instrument operating at an applied magnetic field of 6.5 mT (65 G). Two- and three-dimensional coronal lung images in the anterio-posterior view were obtained during $^3$He breath-holds. 2D images were obtained without slice selection, while 3D images yielded up to eight image planes with a thickness of ~ 1.5 cm. In plane image resolution was ~ 4 mm. Peak SNR is high, being ~ 100 for the 2D projection images, and above 30 for the planes of the 3D datasets. The images show differences in lung shape and size as a function of subject posture, which indicates that the open-access imager will enable posture-dependent pulmonary functional imaging and thereby serve as a valuable tool for the study of critical pulmonary diseases and questions relating to posture-dependent and gravitational effects on pulmonary function. In addition, the open-access imager, operating such a low applied field, could provide lung imaging for subjects with implants, prostheses, claustrophobia or acute illnesses who have been denied access to MRI in its traditional form.

# ACKNOWLEDGEMENTS

Support is acknowledged from NASA grant NAG9-1489, NSF grant CTS-0310006, NIH grant R21 EB006475-01A1, and Harvard University. We thank Dr. Mirko Hrovat, Dr. Jim Maddox, Ms. Rachel Burke, Ms. Ana Batrachenko, Ms. Rachel Scheidegger and Mr. Dan Chonde for technical assistance with imager, and Dr. Michael Barlow for assistance with the novel laser source. We are indebted to Kenneth Tsai, MD, who acted as observing physician for human imaging trials, and George Topulos, MD, who devised the human protocols.

# FIGURE CAPTIONS

**Figure 1**. Photographs of the open-access human MRI system. a) The open-access imaging area, which allows reorientation of a subject. The gap between the two coils is 90 cm, with over 2 m of open space in the other two dimensions. The photograph shows the pair of main $B_0$ coils on their aluminum support flanges, with the gradient coils located parallel to each $B_0$ coil on additional supports bolted to the flanges. b) The entire imager on its customized aluminum framework, located inside an RF-shielded room. Access to the imaging region from outside the room is straightforward.

**Figure 2**. Subjects in the open-access human MRI system. a) Subject on the support table, ready for imaging in the supine position. The $B_1$ coil is slid into position with the aid of positional guides on the table, below the subject support bed. b) Subject sitting on a wooden chair, ready for vertical orientation imaging. The $B_1$ coil is raised and lowered with a wooden support mechanism that allows easy positioning of the subject and ensures the coil returns to the correct position, independent of the subject.

**Figure 3**. Two-dimensional projection $^3$He MR images of human lungs, obtained using the open-access human MRI system, with subjects positioned as shown in Figure 2. a) Image obtained while the subject was lying horizontally, in a supine orientation. b) Image acquired while the subject was sitting vertically. Both images visualize the lungs as if looking at the subject from the front – i.e., the subject's right lung lobe is on the left of the image. Imaging parameters: $B_0$ = 6.5 mT, Larmor frequency = 210 kHz, FOV = 50 cm, NEX = 1, flip angle = 5$^\circ$, TE/TR = 28.5/85.8 ms. Data size = 128 × 64, zero-filled to 128 × 128, total scan time ~ 4 s. This data originally reported in [69].

**Figure 4**. Three-dimensional $^3$He MR image series of human lungs, obtained using the open-access human MRI system, with subject lying horizontally in a supine orientation. All planes visualize the lungs as if looking at the subject from the front – i.e., the subject's right lung lobe is on the left of the image. Image planes represent slices ~ 1.5 cm thick, and progress from anterior (# 1) to posterior (# 8) through that dataset. Imaging parameters: $B_0$ = 6.5 mT, Larmor frequency = 210 kHz, FOV = 50 × 50 × 12 cm, NEX = 1, flip angle = 4°, TE/TR = 28.5/85.8 ms. Data size = 128 × 64 × 6, zero-filled to 128 × 128 × 8, total scan time ~ 30 s.

**Figure 5**. Three-dimensional $^3$He MR image series of human lungs, obtained using the open-access human MRI system, with subject positioned vertically. Additional room air was not inhaled following $^3$He inhalation, resulting in non-uniform $^3$He distribution throughout the lung, and intense signal in the trachea and oral cavity. MR signal below the diaphragm in each image, beside the plane number, is most likely due to gas above the trachea and outside the top of the image field-of-view that was folded in to the bottom portion of the image. Image orientation, layout and acquisition parameters are the same as for Figure 4.

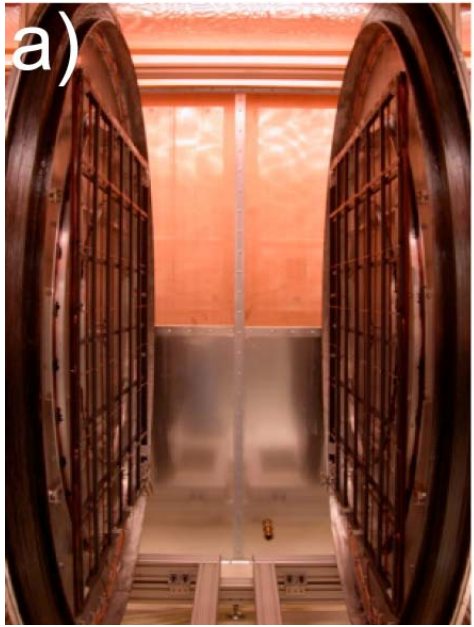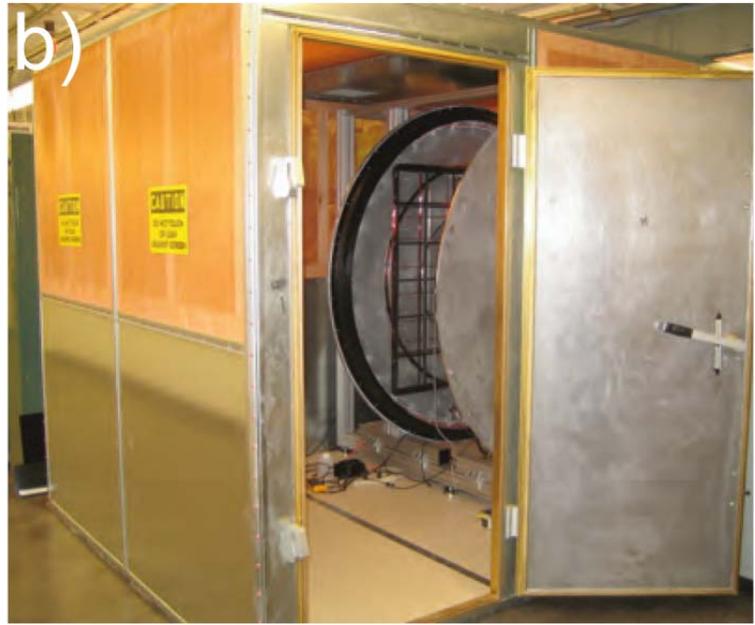

**Figure 1**

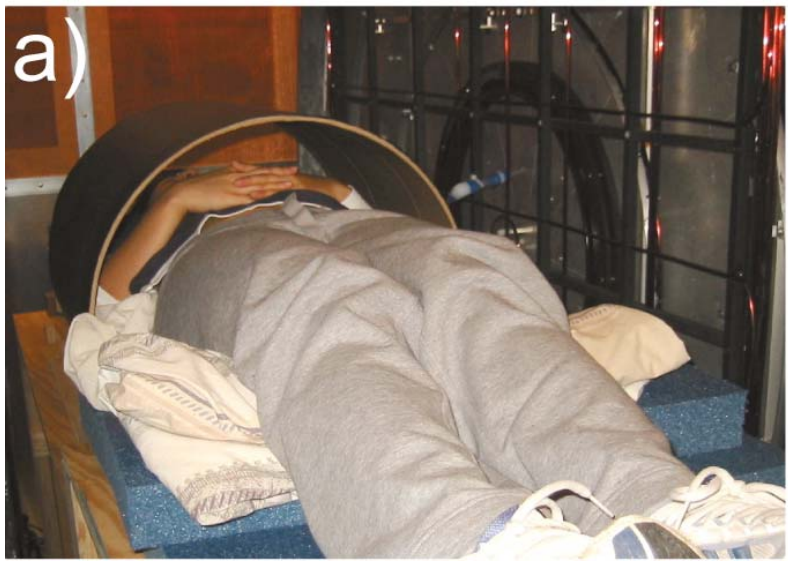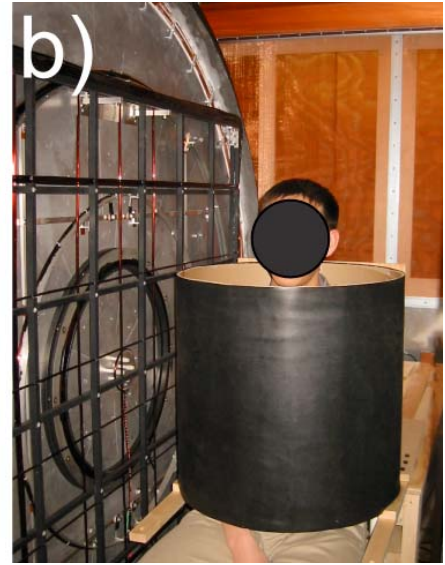

**Figure 2**

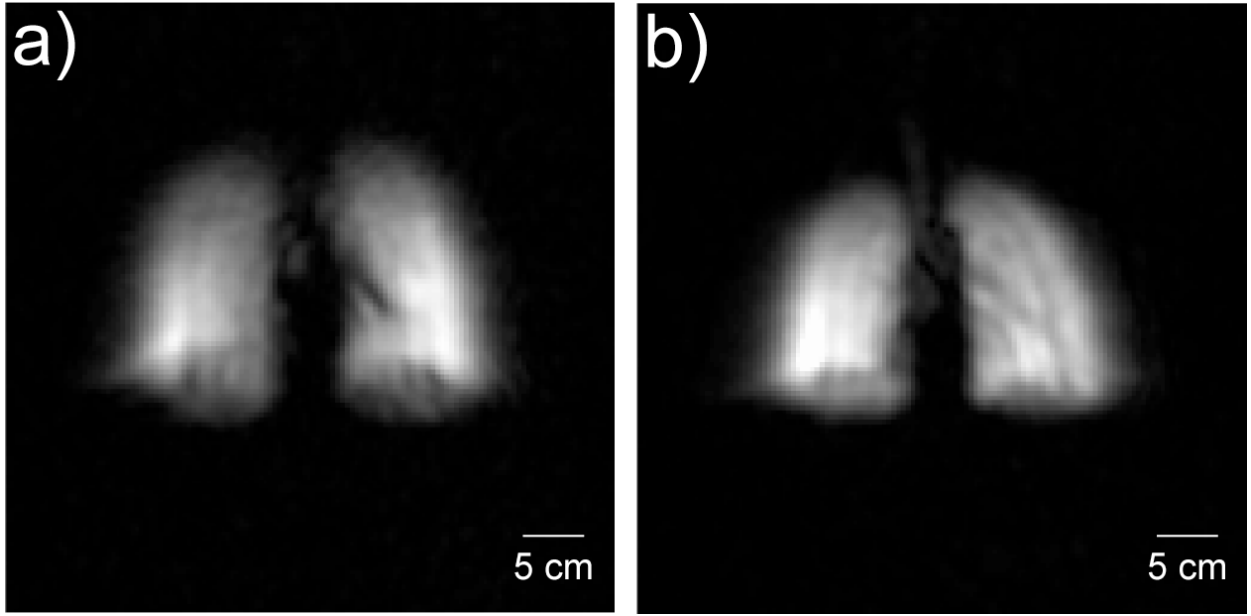

**Figure 3**

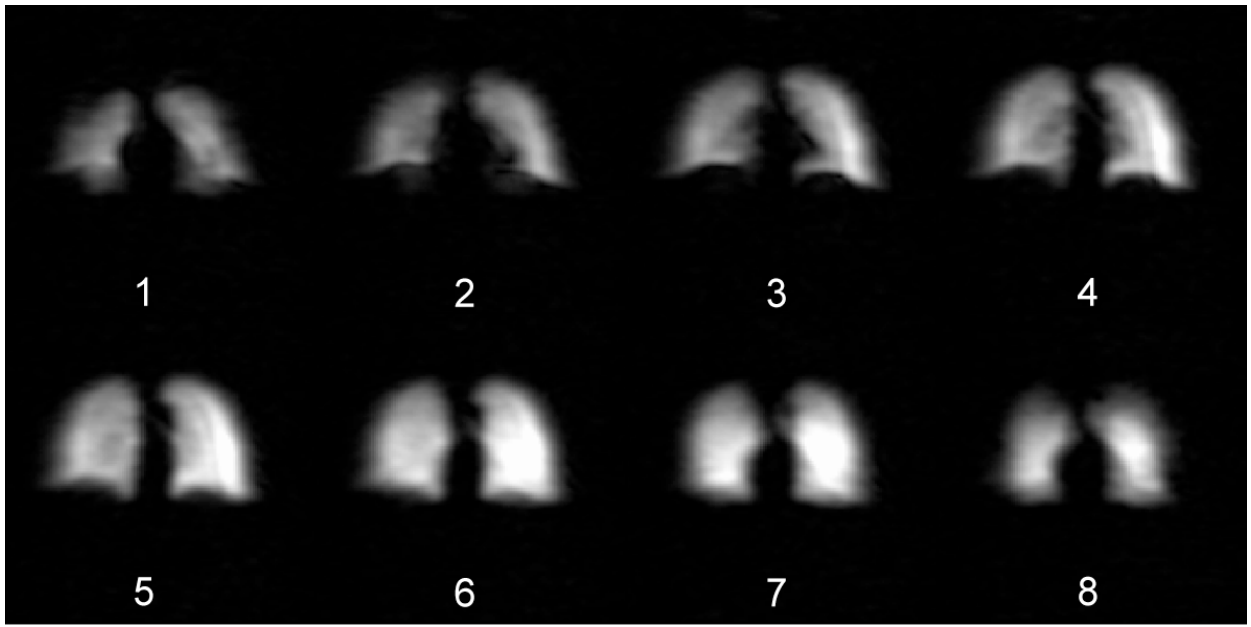

**Figure 4**

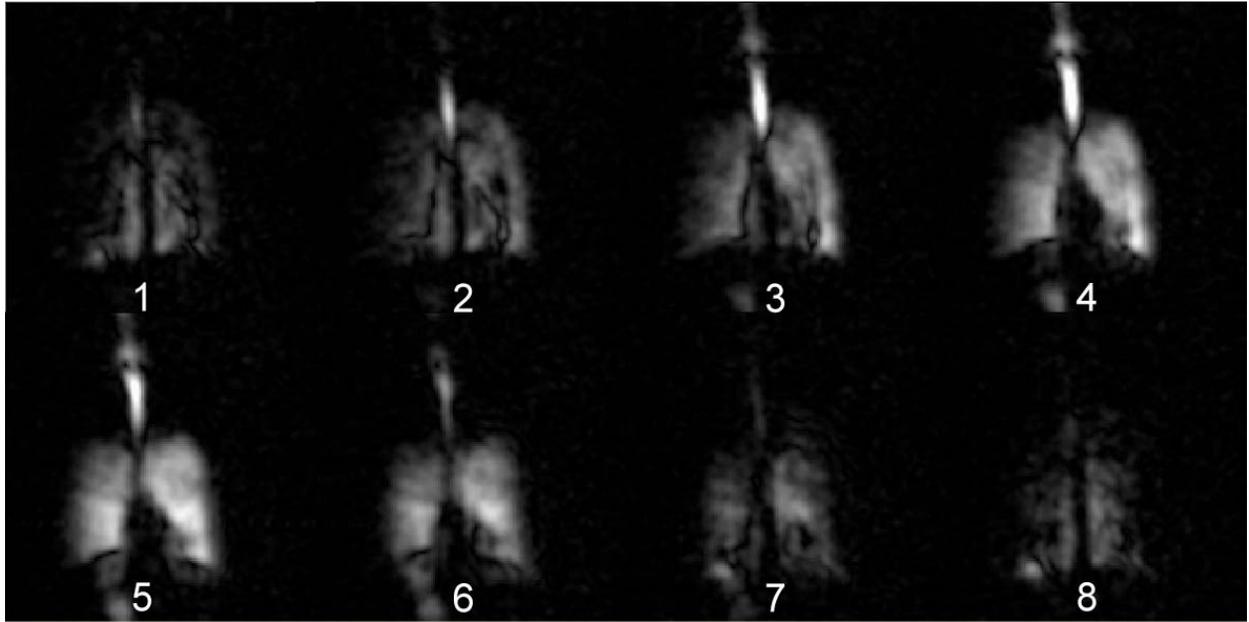

**Figure 5**